# MacroQueue: Automating Measurements in High-Dimensional Parameter Spaces

*Brad M. Goff and Jay A. Gupta\**
*Department of Physics, The Ohio State University, Columbus, OH 43210, USA*
*\* Corresponding author: gupta.208@osu.edu*

**Abstract**

Laboratory measurements often use several instruments to fully explore the relavent parameter space; such as, an external lock-in amplifier, an electromagnet, a RF generator, etc.. Ordinarily, these instruments must be individually controlled, and their parameters must be manually recorded. MacroQueue enables effortless measurements throughout these high-dimensional parameter space systems. MacroQueue is a modular software designed for controlling and automating various laboratory equipment in sync without requiring significant coding ability. It can be extended to automate any system that can be controlled via Python.

Users can easily add python functions to control new and existing auxiliary equipment, in addition to new and existing systems, if they can be controlled via python. Although any arbitary python function can be added, it is preferable that functions are written with the functional programming paradigm in mind, so they are small, and each perform a single task. This allows the functions to be reused for many types of measurements.

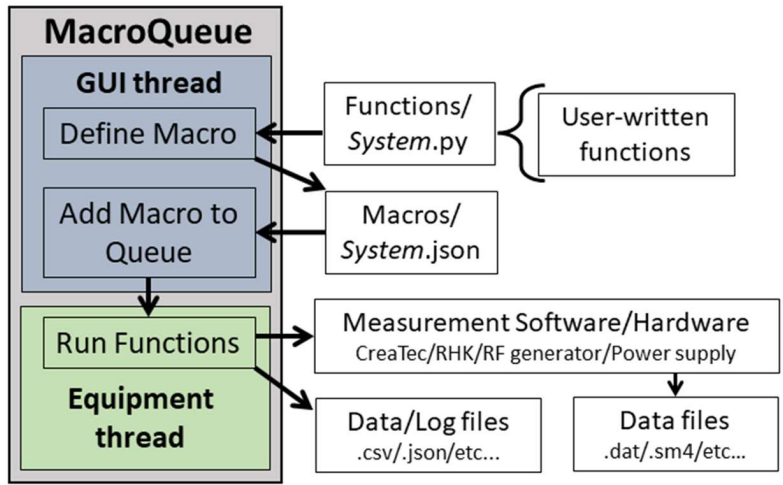

*Figure 1*: The general procedure for using MacroQueue.

The functions are grouped into a macro for each type of measurement. Macros are added to a queue with different values for each parameter, e.g. bias, magnetic field, etc., to perform measurements throughout the parameter space. Each measurement is performed consecutively on a separate thread to allow measurements that are still in the queue to be modified.

These features allow users to easily control several instruments in sync, add new instruments to a system, and perform long series of measurements with minimal input.

## Statement of need

Numerous instruments must be controlled in sync to access the full parameter space of a system. MacroQueue provides a simple GUI to allow users to perform measurements throughout the parameter space without coding. This software is currently in active use in several laboratories at The Ohio State University and the NSF NeXUS Facility. It assisted in the research of @Goff_2024 and @koll2024formation. Several similar packages and APIs already exist, such as @PyMeasure, @bluesky, and @ScopeFoundry2023. A good overview of available Python packages can be found in @buchner_2022_6399528. The goal of MacroQueue is to provide a frontend GUI that allows users to perform measurements in high-dimensional parameter spaces without requiring the coding ability that is necessary to use the existing APIs while still providing advanced users the flexibility to write arbitrarily complex functions.

## Overview

MacroQueue can be packaged into an executable, by either using the provided script or using PyInstaller directly, so that coding and using the command line is completely optional. To further allow the user experience to be as simple as possible, any arbitrary python function, no matter how basic, can be added to MacroQueue. Even as an executable, users can open the "source folder" via the File menu to access the python files that control various instruments. Upon launching, MacroQueue searches this folder for new files. Every function, from each python file, is dynamically imported using the package importlib, part of python's standard library. For each parameter in a function, MacroQueue reads the default value to interpret the datatype (e.g. string, numerical, boolean, list, file path) and the appropriate control in the GUI (e.g. text box, numbers only text box, checkmark, dropdown menu, file browser respectively). Figure 2 shows example code and the various controls that are produced in the GUI.

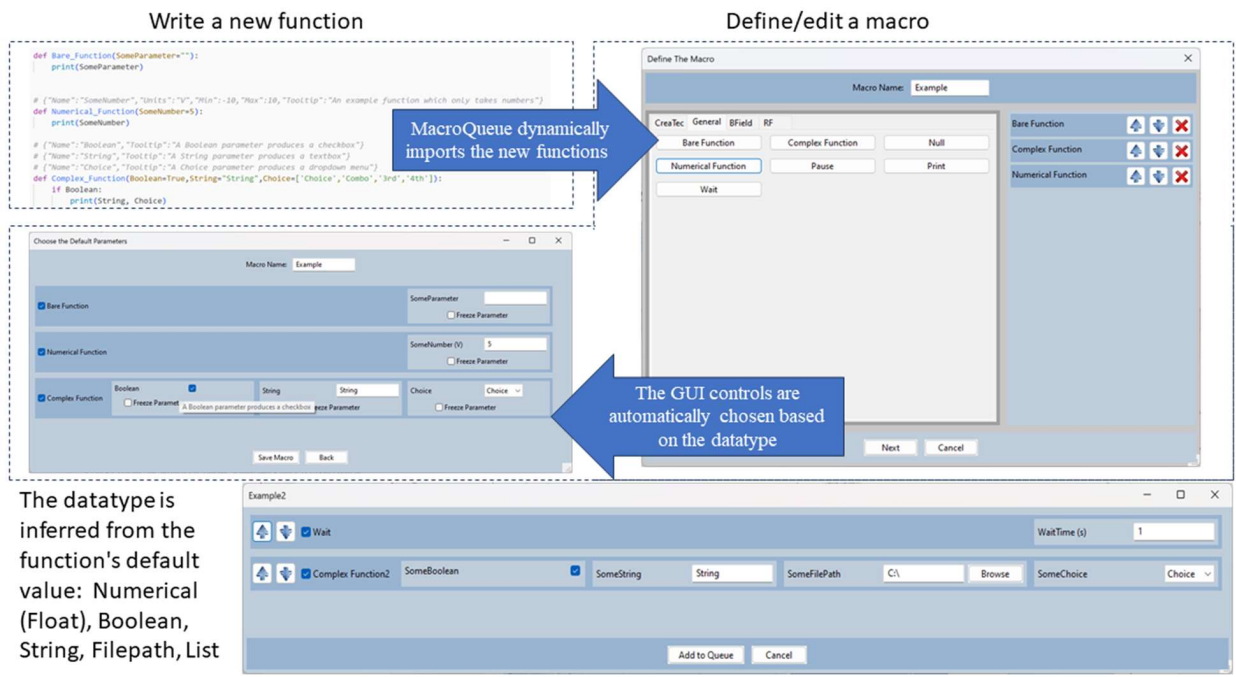

*Figure 2*: The workflow for adding a new function and defining a new macro

For additional features, users can write metadata for each parameter in comments above each function. The metadata includes units and an explanation of what the parameter does, which will be included in the parameter's hover tooltip. Numerical values can also have a soft minimum and/or maximum value in their metadata; when a users tries to input a value outside the range, there will be a pop-up warning to confirm that they want to proceed. Hard limits can be applied in individual function by raising an exception, such as a ValueError exception. If an exception is thrown in any of the functions, intentional or otherwise, the queue will be paused, the current macro will be canceled, and a pop-up will provide the user with the exception's details.

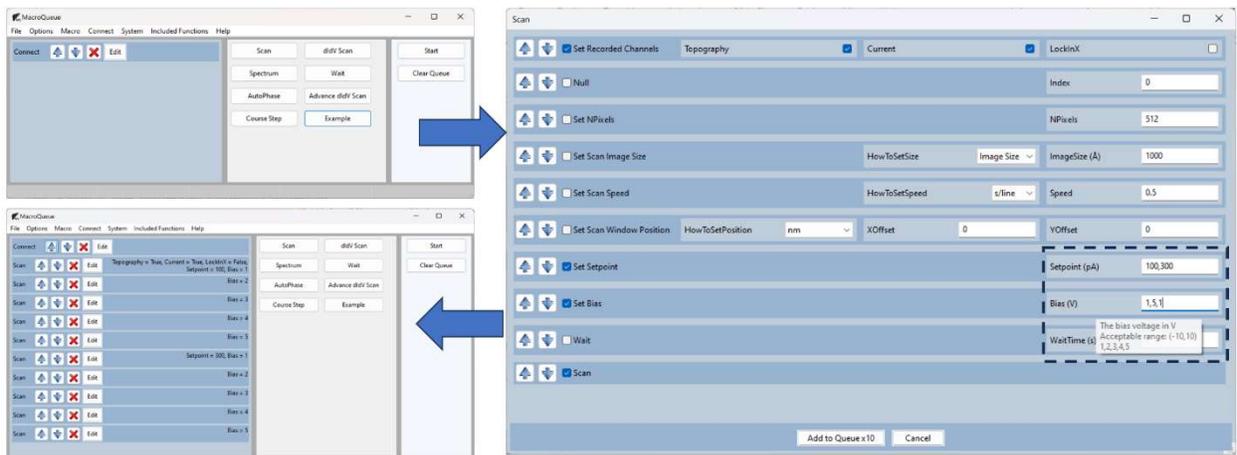

*Figure 3*: The workflow for adding macros to the queue.

MacroQueue makes a GUI, shown Figure 3, using the WxPython toolkit @WxPython.
The queue is on the left, showing the macros that will be run. The existing macros are on the right. New macros can be created via the Macro menu or by modifying an existing Macro. When you add a macro to the queue, there will be a menu where you can edit which functions in the macro will be run and the values for each parameter that will be used. Users can input multiple values for numerical parameters. MacroQueue will 'expand' the macro into multiple macros, with each value, to the queue. The macros are created as if there is a for loop at the expanding function and every function below the expanding function is inside the loop. Figure 3 shows an example with 2 values for the setpoint and 5 values for the bias so 10 macros are created. The setpoint function will run twice, and the bias function will run 10 times, 5 times per setpoint. This can be used to quickly add thousands of measurements throughout the parameter space to the queue.

**Acknowledgements**

This work was primarily supported by the Department of Energy (DOE) Basic
Energy Sciences under Grant No. DE-SC0016379.